\documentclass[aps,prl,amsmath,showpacs,twocolumn]{revtex4}

\usepackage[latin1]{inputenc}
\usepackage[T1]{fontenc}
\usepackage[dvips]{graphicx}
\usepackage{color}
\usepackage[UKenglish]{babel}
\usepackage{subfigure}
\usepackage{ulem}

\newcommand{\VERT}{\mathchoice{\big|}{\big|}{|}{|}}
\newcommand{\RANGLE}%
{\mathchoice{\bigr\rangle}{\bigr\rangle}{\rangle}{\rangle}}
\newcommand{\LANGLE}%
{\mathchoice{\bigl\langle}{\bigl\langle}{\langle}{\langle}}

\newcommand{\braket}[2]{\LANGLE{#1}\VERT{#2}\RANGLE}

\begin{document}

\title{Attosecond control of electron dynamics in carbon monoxide}

\author{I. Znakovskaya$^1$, P. von den Hoff$^2$, S. Zherebtsov$^1$, A. Wirth$^1$, O. Herrwerth$^1$, M.J.J. Vrakking$^3$, R. de Vivie-Riedle$^2$, M.F. Kling$^1$}

\affiliation{$^1$Max-Planck Institute of Quantum Optics, Hans-Kopfermann-Str. 1, 85748 Garching, Germany\\$^2$Department für Chemie und Biochemie, Ludwig-Maximilians-Universit\"at M\"unchen, D-81377 M\"unchen, Germany\\$^3$FOM-Institute AMOLF, Science Park 113, 1098 XG Amsterdam, The Netherlands\\Emails: regina.de\_vivie@cup.uni-muenchen.de; matthias.kling@mpq.mpg.de}

\begin{abstract}
Laser pulses with stable electric field waveforms establish the opportunity to achieve coherent control on attosecond timescales. We present experimental and theoretical results on the steering of electronic motion in a multi-electron system. A very high degree of light-waveform control over the directional emission of C$^+$ and O$^+$ fragments from the dissociative ionization of CO was observed. Ab initio based model calculations reveal contributions to the control related to the ionization and laser-induced population transfer between excited electronic states of CO$^+$ during dissociation.
\end{abstract}

\maketitle

\begin{figure}[htbp]
  \centering
  \subfigure{
    \includegraphics[trim=0 10 0 10,width=0.57\columnwidth]{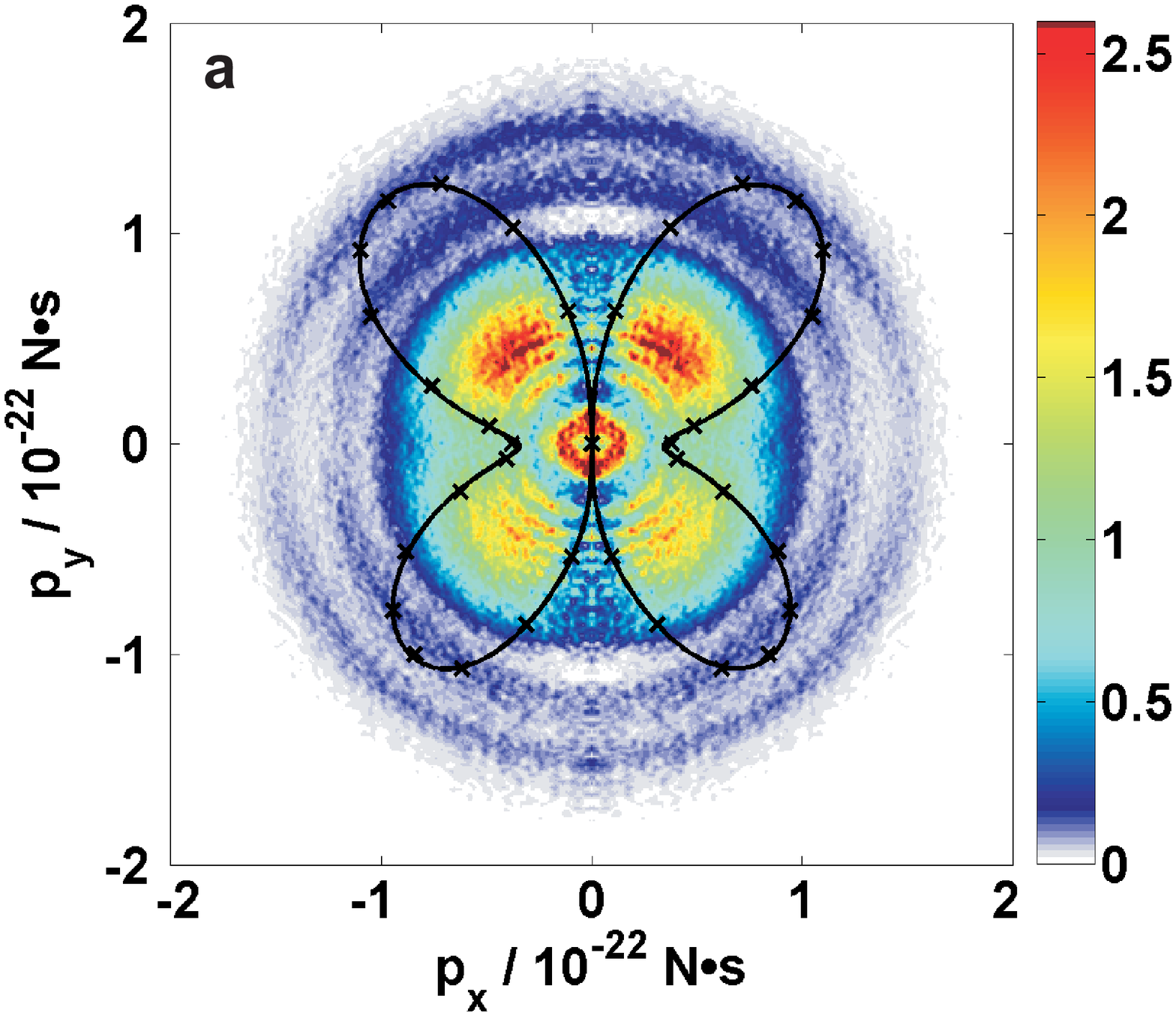}
}
  \subfigure{
    \includegraphics[trim=120 60 120 15,clip,scale=0.30]{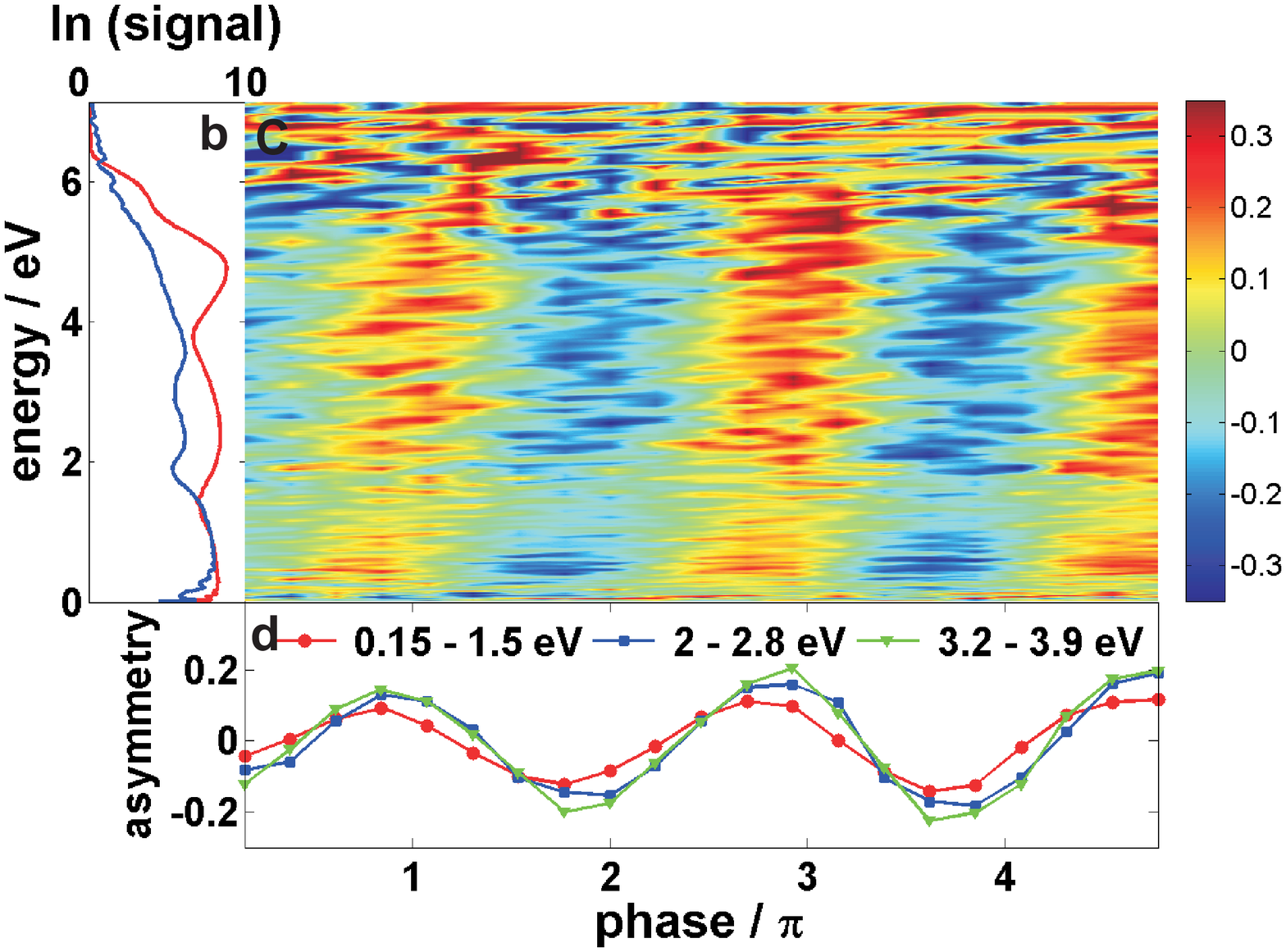}

}
  \caption{(a) Inverted two-dimensional C$^+$ momentum image (the laser polarization is vertical) and calculated ionization rate of CO in arb. un. (black line) for CEP$\simeq\pi$; (b) measured (blue) and theoretical (red) C$^+$ kinetic energy spectrum for CO dissociative ionization. (c) Asymmetry of C$^+$ ion emission along the laser polarization (integrated over 120°) vs. kinetic energy and phase. (d) Asymmetry integrated over indicated energy ranges vs. CEP. The CEP was calibrated by reference measurements in Xe and comparison to calculations based on the quantitative rescattering theory according to a recently published method \cite{micheau09_short}. From the evaluation of this data we obtain an error in the absolute CEP of $\pm$0.04$\pi$.}
  \label{Fig:Exp_data}
\end{figure}

\begin{figure}
 \centering
 \includegraphics[trim=0 15 0 0,width=0.85\columnwidth]{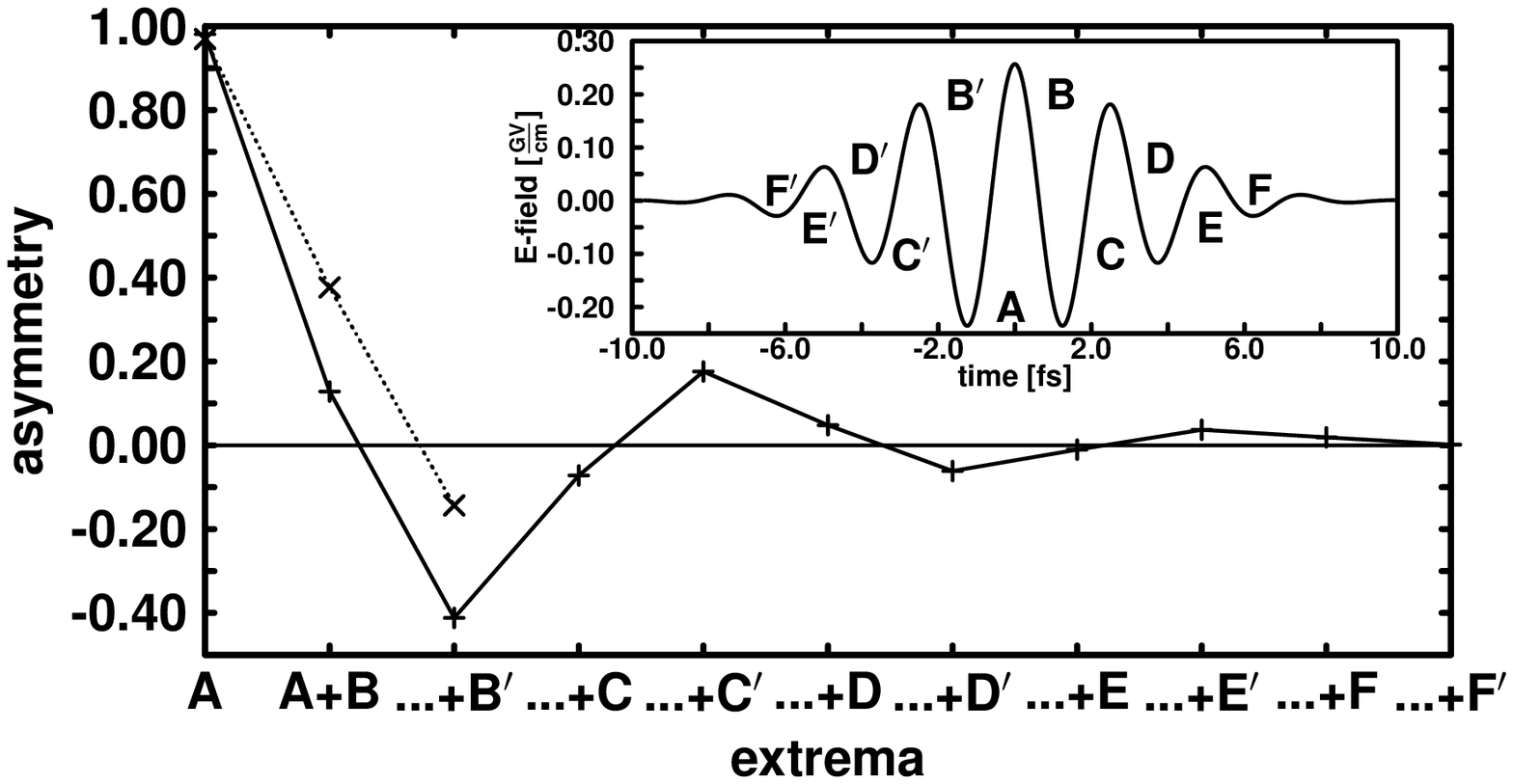}
 \caption{Calculated asymmetries (+ symbols and solid line) for various combinations of half-cycles of the laser field (shown in the inset). The x symbols show volume averaged asymmetry contributions (dashed line).}
 \label{fig:asym}
\end{figure}

Coherent control of chemical reactions and photobiological processes has been achieved by manipulating the laser frequency, phase and polarization in closed loop experiments~\cite{brixner03_short}.
Control of the electric field waveform $E(t)=E_0(t)\cos(\omega t+\phi)$, with envelope $E_0(t)$, and frequency $\omega$, by the carrier envelope phase (CEP) $\phi$ constitutes a new paradigm of coherent control. This control became accessible with CEP-stabilization and opened the door to steer electrons in atomic and molecular systems on attosecond timescales~\cite{kienberger07_short}. Waveform controlled few-cycle pulses have only recently been used to control electron localization in the dissociative ionization of the prototype molecules D$_2$ ~\cite{kling06_short,kling08a_short} and HD~\cite{kling08a_short}. The same processes were also theoretically studied (see e.g.~\cite{Ben-Itzhak04_short,graefe07,tongLin07,Geppert2008_short,bandrauk04_short} and references cited therein). After initial ionization, these systems only contain a single electron.
The important question arises, whether the steering of electrons in more complex systems is feasible and --- if yes --- can we understand the role of the initial ionization/excitation process and the following strong-field coupling of the various potential energy surfaces in the observed control?
We describe experiments, where control of electron dynamics in carbon monoxide (CO) was achieved by the light-waveform. Phase-stabilized 4~fs, linearly polarized laser pulses at 740~nm and at an intensity of 8$\,\times\,$10$^{13}$~W~cm$^{-2}$ were applied to dissociatively ionize CO. The directional emission of ionic fragments was monitored via velocity-map imaging (VMI). We compare the experimental results to full quantum calculations allowing a mechanistic interpretation and understanding of the observed control.
The dynamics of molecules in strong laser fields typically includes ionization and dissociation. While there is a wealth of work on the dissociation of small molecules in strong laser fields (see e.g.~\cite{posthumus04} and references cited therein), only a limited number of studies have been performed on CO. Guo studied the multiphoton induced dissociative ionization of CO at 800~nm and an intensity of 3.8$\,\times\,$10$^{13}$~W~cm$^{-2}$~\cite{guo06}. The yield of C$^+$ ions from the dissociation was found to be more than an order of magnitude higher than that of O$^+$ fragments. Experiments reported by Alnaser et al.~\cite{alnaser05_short} using 8\,fs pulses at 800\,nm and an intensity of 6$\,\times\,$10$^{13}$~W~cm$^{-2}$ showed an angular distribution of the ionic fragments from the Coulomb explosion of CO with a maximum along the laser polarization axis, for which the shape of the highest occupied molecular orbital (HOMO) of the CO molecule was held to be responsible.
\begin{figure}
 \centering
 \includegraphics[trim=0 30 0 0 ,width=0.8\columnwidth]{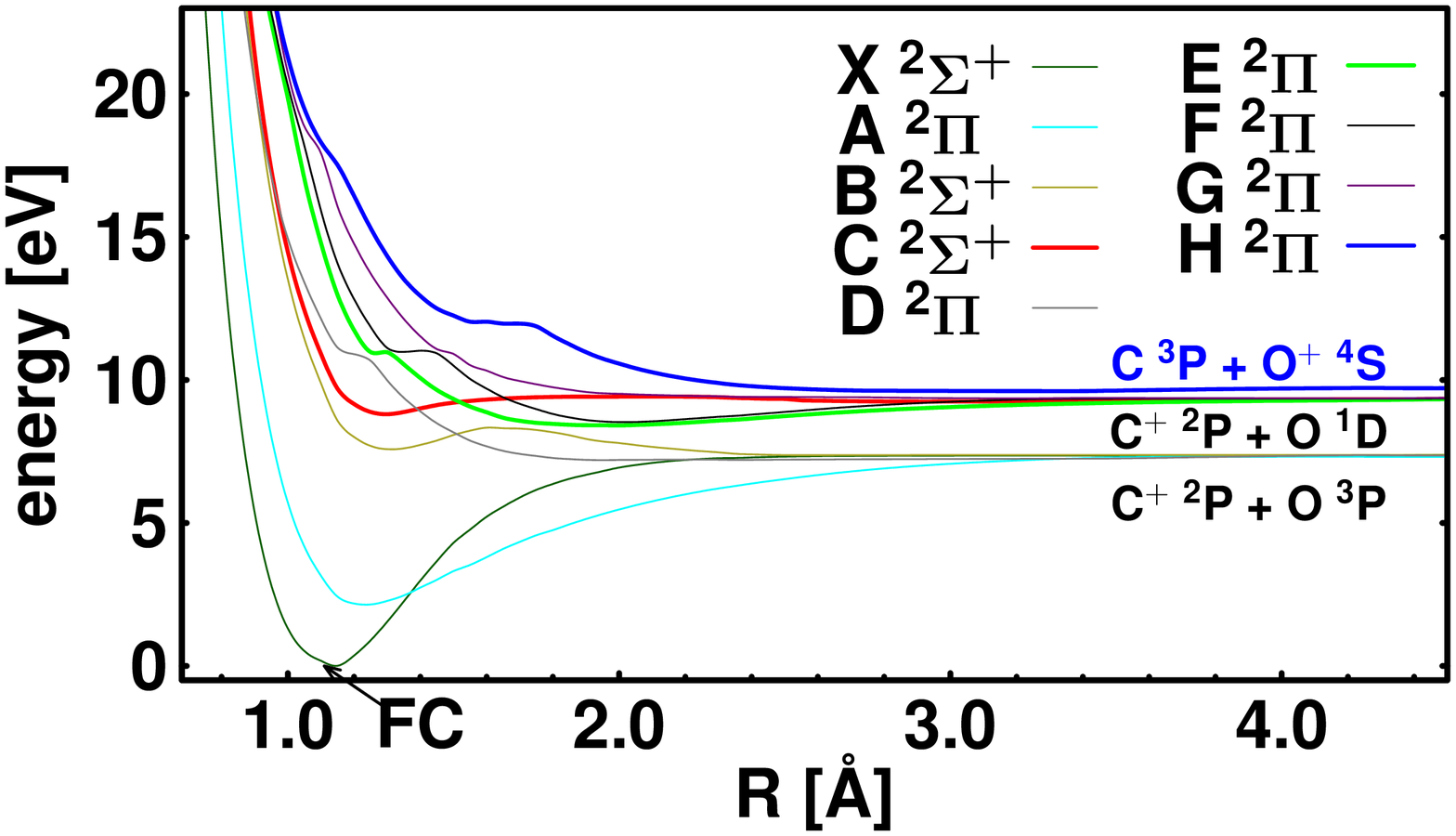}
 \caption{PES of CO$^+$ and Frank Condon point (FC) for the ionization from neutral CO obtained by calculations described in the text.}
 \label{fig:Pot}
\end{figure}
Here, we report on the CEP-dependent directional emission of ionic fragments from the dissociative ionization of CO. To demonstrate the complexity of the molecular system, we show in Fig.~\ref{fig:Pot} the calculated potential energy surfaces (PES) of the lowest $^2\Sigma^+$- and $^2\Pi$-states relevant for the photodissociation of CO$^+$ at our laser intensities. In the initial ionization, the low-lying bound electronic states X and A of CO$^+$ can be populated. Higher electronic states of CO$^+$ can be reached by recollision or multi-photon excitation starting the dissociation process. At our intensity the recollision energy of the electron from the initial ionization is up to 13~eV.
The setup used in these experiments was described earlier~\cite{schultze07_short}. Briefly, phase-stabilized few-cycle pulses were focused into the center of the ion optics of a VMI spectrometer  using a spherical mirror (f~=~125~mm). Ions and electrons that were generated at the crossing point of the laser (linearly polarized along the $y$-axis and propagating along the $x$-axis) and an effusive atomic/molecular beam were projected (along the $z$-axis) onto a multi-channel plate (MCP)/phosphor screen assembly and recorded with a cooled CCD camera. Inversion of the recorded projections using an iterative procedure allowed reconstruction of the original 3D ion momentum distributions.
Fig.~\ref{Fig:Exp_data}(a) shows a cut through the 3D momentum distribution in the $xy$-plane at $p_{z}$=0 for C$^+$ ions and for a CEP$\simeq\pi$ pulse. An up-down (positive $p_y$ versus negative $p_y$ values) asymmetry in the C$^+$ fragment emission is visible and the contributions may be best identified as three rings, where the first ring is the broadest (from $p$ = 0 to 1$\cdot 10^{-22}Ns$) and most intense and exhibits additional sharp lines. The other rings appear between momenta of 1.1 to 1.3 and 1.4 to 1.6$\cdot 10^{-22}Ns$. The dominant features show angular distributions with minima around 0°, 90°, 180° and 270° with varying modulation depths. This angular distribution of the C$^+$ fragments differs from the findings in ref.~\cite{alnaser05_short} and indicates contributions from the HOMO (5$\sigma$) and the HOMO-1 (1$\pi_{x/y}$). The solid black line in Fig.~\ref{Fig:Exp_data}(a) displays the calculated volume-averaged angle-dependent ionization rate as described below in the theory section.
Fig.~\ref{Fig:Exp_data}(b) shows the kinetic energy spectrum derived from Fig.~\ref{Fig:Exp_data}(a) by angular integration (blue line) together with the calculated spectrum obtained within the theoretical approach described below (red line). Comparison with experiments using circularly polarized light at twice the intensity as for linear polarization, which suppresses the recollision excitation, reveals a distinct decrease in intensity of all three rings. Thus, we conclude that recollision excitation is responsible for the production of the observed ionic fragments at our experimental conditions.
In the fragmentation of CO$^+$, the C$^+$+O channel is energetically favored over the C+O$^+$ channel (see Fig.~\ref{fig:Pot}). We find the C$^+$ yield to be approx. 20 times larger than the O$^+$ yield. Throughout the measured CEP range a pronounced phase dependence in the directional ion emission is found for C$^+$ as well as for O$^+$ ions. The directional emission is represented by the asymmetry $A(W,\phi)= \frac{P_{\mathrm{up}}(W,\phi)-P_{\mathrm{down}}(W,\phi)}{P_{\mathrm{up}}(W,\phi)+P_{\mathrm{down}}(W,\phi)}$ as a function of the fragment kinetic energy $W$ and the laser phase $\phi$.
$P_{\mathrm{up}}(W,\phi)$ and $P_{\mathrm{down}}(W,\phi)$ are the angle-integrated ion yields in the up and down directions. Fig.~\ref{Fig:Exp_data}(c) displays the observed asymmetry $A$($W,\phi$) for the dissociative ionization of CO into C$^+$ and O as a function of the CEP and the kinetic energy $W$ of the C$^+$ ion fragments. Fig.~\ref{Fig:Exp_data}(d) shows the asymmetry parameter integrated over selected energy ranges. The observed asymmetry in the directional emission of C$^+$ ions is very pronounced and almost equally strong throughout the kinetic energy spectrum. 
A similar asymmetry map (not shown here) was recorded for O$^+$ ions, showing the same features. Due to the significantly weaker O$^+$ signal, however, the asymmetry map exhibits a lower signal-to-noise ratio, making it also difficult to determine the phase-shift between the points, where a maximum asymmetry is found for C$^+$ and O$^+$ ions.
In the experiment, the asymmetry can arise from contributions of all three steps: ionization,  recollisional excitation and laser-induced population transfer between excited electronic states of CO$^+$. The initial population of excited states by recollision is likely dependent on the CEP, but its calculation is currently out of scope for larger molecules~\cite{graefe07}. We can, however, focus on the remaining two steps. The ionization probability of a molecule in a laser field is determined by the electron flux induced by this external electric field~\cite{Smirnov1966}. To calculate the angular dependent ionization probability for a given static electric field, we performed quantum chemical calculations under various orientation angles with respect to the polarization of the applied external field. In the spirit of~\cite{Smirnov1966} we record the electron flux through a surface placed at the outer turningpoints of a given orbital. When ionization from the HOMO only is considered as in ref.~\cite{alnaser05_short}, our calculated angular distribution matches the reported results therein. For our experimental conditions we allow ionization from the HOMO (leading to the X $^2\Sigma^+$ state of CO$^+$) and the HOMO-1 (leading to the A $^2\Pi$-state) reproducing the observed angular distribution (Fig.~\ref{Fig:Exp_data}(a)). This is achieved by a basis transformation forming the orbitals HOMO+HOMO-1 and HOMO-HOMO-1. By integration of the calculated ionization probabilities over the appropriate orientation angles, the asymmetry in the ionization step caused by a phase-stabilized electric laser field can be extracted. Assuming ionization only at the extrema of the field, we calculated the asymmetries for various contributions of half-cycles as shown in Fig.~\ref{fig:asym} (solid line). The volume-averaged angular dependent ionization probability is obtained for a spatial gaussian intensity profile and is shown in Fig.~\ref{Fig:Exp_data}(a) (black line) for the combination of the three most prominent half-cycles and a CEP=$\pi$ pulse. The related volume-averaged asymmetries are presented in Fig.~\ref{fig:asym} (dashed line). The remaining less intense half-cycles can be neglected as they do not cause recollision excitation. The good agreement between theory and experiment shows that the angular distribution arises from the ionization out of two orbitals. Its asymmetry originates from phase stable electric fields. The corresponding calculated asymmetry amplitude of 0.14 is below the experimentally observed total asymmetry amplitude of 0.2. If only ionization would be responsible for the observed CEP-dependence, the asymmetry would be expected to be seen for all observed C$^+$ ions irrespective of their kinetic energy or angular distribution. Experimentally, we do however find that the angular range over which the asymmetry is observed depends on the fragment kinetic energy, suggesting that the experimental result is not solely explained by the ionization mechanism.
To calculate to what extent the asymmetry may arise in the dissociation process we used our recently introduced approach to describe coupled electron and nuclear dynamics~\cite{Geppert2008_short}. We follow the dominant recollision excitation pathway, assuming ionization of CO only at the electric field maximum of the laser pulse used in the experiment and recollision of the electron after 1.7~fs~\cite{niikura02_short}. Our quantum dynamical simulations start after the recollision.
The initially formed nuclear wave packet is excited from the X and A state to higher electronic states. Thereby a coherent superposition of several electronic states is created and simultaneously the electronic and nuclear wavepacket motion is initiated.
The PES for the CO electronic ground state and for the ionic states are calculated with \textsc{Molpro}~\cite{MOLPRO_short}. 
From the set of CO$^+$ PES we chose three potentials as typical representatives for the induced dynamics which moreover allow all transitions among each other. We include the C~$^2\Sigma^+$ state as the weakly bound state and the E~$^2\Pi$ state to resemble the repulsive dynamics.
Both states correlate with the C$^+$($^2$P)+O($^1$D) channel. As third PES we include the H~$^2\Pi$ state which is the first state leading to the C($^3$P)+O$^+$($^4$S) channel, delivering O$^+$ fragments as observed in the experiment. The initial wavepacket is composed as a 55:38:7 distribution of the states involved, ordered in increasing energy, assuming a gaussian energy distribution for the recolliding electron with a cut off energy of 13~eV. Choosing these particular values for the initial populations also reflects the experimentally observed ratio between C$^+$ and O$^+$ fragments. In the calculations the CO$^+$ ions are taken to be aligned at 45$^{\circ}$ to the laser polarization as the ionization peaks along this direction and all transitions between $\Sigma$ and $\Pi$ states are allowed.
\begin{figure}
 \centering
 \includegraphics[trim=0 15 0 0,width=0.85\columnwidth]{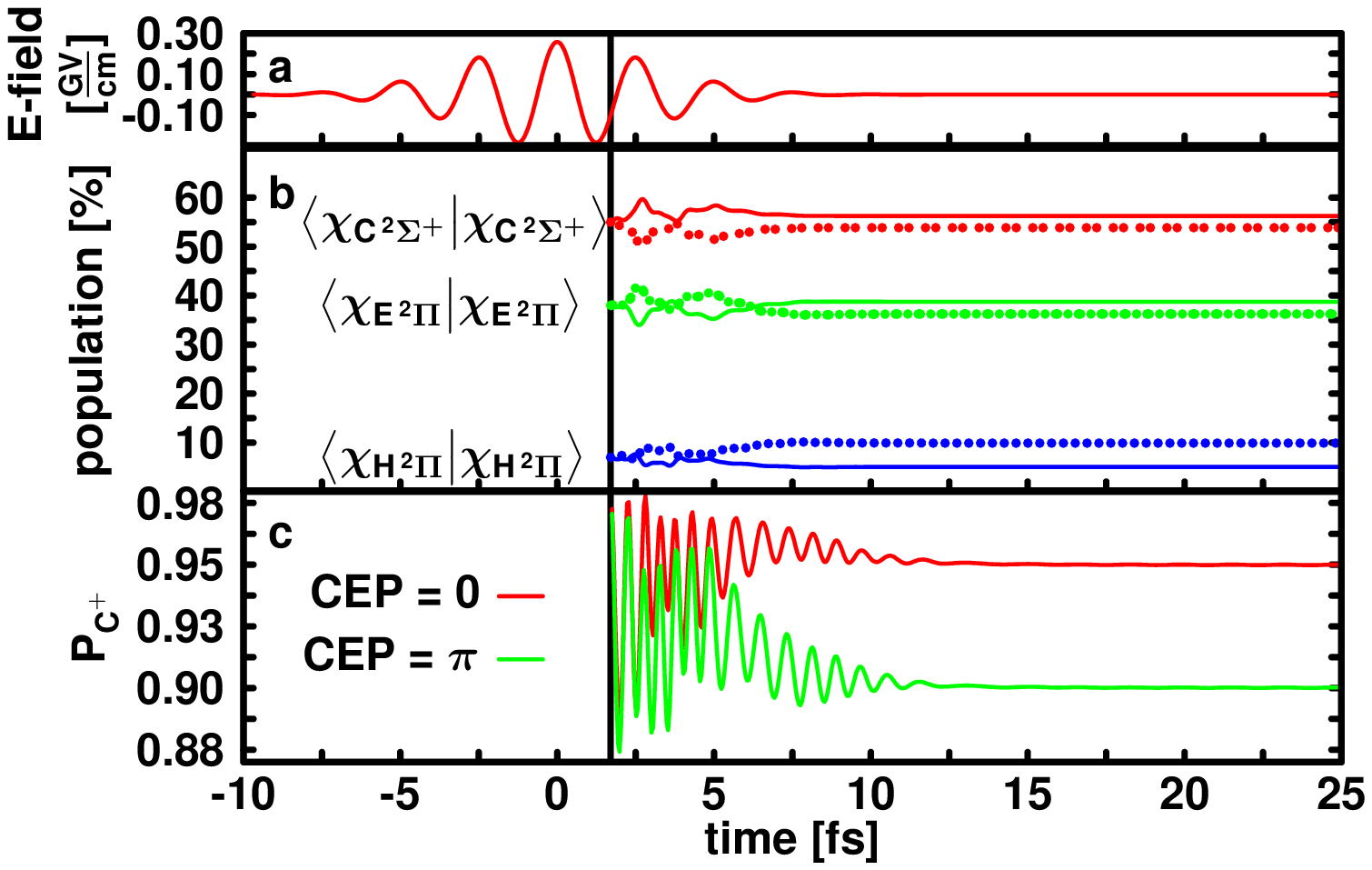}
  \caption{(a) Electric field (CEP~=~$\varphi$=0). (b) Time-dependent populations of the C $^2\Sigma^+$, E $^2\Pi$ and H $^2\Pi$ states after recollision excitation (solid: CEP=0; dotted: CEP=$\pi$). (c) Temporal evolution of the probability measuring a C$^+$ fragment $P_\mathrm{C^+}$.}
 \label{fig:QD}
\end{figure}
The molecular wavefunction $\Psi_\mathrm{mol}=\sum_{i}\chi_i(R,t)\varphi_i(r,t;R)$ is set up as the sum over these three electronic states $i$
with $\chi_i$ the nuclear wavefunctions, $\varphi_i$ the field-free electronic wavefunctions, the nuclear and electronic coordinates $R$ and $r$ and the time $t$. Figure~\ref{fig:QD} shows the temporal evolution of the laser field (a) and of the population in the three selected electronic states (b). The calculated kinetic energy spectrum (Fig.~\ref{Fig:Exp_data}(b)) derived from the nuclear dynamics is in reasonable qualitative agreement with the experimental data supporting the selection made for the representative states. Note that contributions from doubly charged states cannot be completely ruled out in the experimental spectrum. On the basis of these results, we can explain the origin of the observed energy distribution. The low kinetic energy spectrum arises from the dynamics on the weakly bound state while the spectrum between 2.0\nobreakdash--2.8~eV originates from the purely repulsive state. The high energy spectrum reflects the dynamics of the second repulsive state correlating with the O$^+$ channel. The sharp peaks in the low kinetic energy spectrum, also present in the experimental data (Fig.~\ref{Fig:Exp_data}(a,b)), arise from interference of bound and dissociative vibrational states in the nuclear wave packet on the C $^2\Sigma^+$ state. The electronic density $\rho(r_1,t;R)$ is expressed as a function of the electronic coordinate $r_1$ and time. The total molecular wavefunction is integrated over the nuclear and over the $N-1$ electronic coordinates (with $N$ the total number of electrons):
\begin{equation}
 \begin{split}
 \label{eq:Dens}
 &\rho(r_1,t;R)= \int \Psi^*_{\mathrm{mol}} \Psi_{\mathrm{mol}} \, \textrm dR \, \textrm dr_2 \ldots \textrm dr_{N}\\ &=\sum_{i=1}^3 a_i(t)^2 \VERT \varphi_i(r_1,t_0;R)\VERT^2\\&+ \sum_{i=1}^3 \sum_{j<i}2 Re \Big\{ \braket{\chi_i(R,t)}{\chi_j(R,t)}_{_R} \\&\varphi_i(r_1,t_0;R) \varphi_j(r_1,t_0;R) e^{- \textrm i\Delta E_{ji}(R(t)) \Delta t +\phi(t-\Delta t)} \Big\}
\end{split}
\end{equation}
with $a_i(t)=\sqrt{\braket{\chi_i(t)}{\chi_i(t)}_{_R}}$ and $a_i^2(t)$ the population of the electronic states, the interference term $\braket{\chi_i(t)}{\chi_j(t)}_{_R}$ and the energy difference $\Delta E_{ij}(t)=E_j(t)-E_i(t)$ between the electronic states $i$ and $j$. The electronic wavefunctions $\varphi_i(r,t_0;R)$ are represented as Slater determinants. The time evolution of the electronic wavepacket is calculated by propagation in the eigenstate basis. The coupling of the fast electron to the slower nuclear dynamics enters through the time-dependent population and through the interference term, which specifies the degree of electronic coherence induced in the molecular system.
The probability $P_\mathrm{C^+}(t)$ of measuring a C$^+$ fragment is given by $P_\mathrm{C^+}(t)=\int_{x_\mathrm{min}}^{x_\mathrm{max}}\mathrm{d}x\int_{y_\mathrm{min}}^{y_\mathrm{max}}\mathrm{d}y\int_{z_\mathrm{min}}^0\mathrm{d}z \,\rho(r_1,t;R(t))$ where $x,y$ and $z$ refer to the molecular frame with $z$ along the molecular axis and the O-atom oriented along negative $z$-values~\cite{Geppert2008_short}. In the first 6~fs the electron dynamics, reflected in $P_\mathrm{C^+}(t)$ (e.g. red curve in fig.~\ref{fig:QD}(c)), results from a competition between the influence of the light pulse, the dynamics of the linear combination and the interference term of the nuclear wavefunctions. As soon as the population transfer between the electronic states stops (approximately after 8~fs, see fig.~\ref{fig:QD}(b)) $P_\mathrm{C^+}(t)$ oscillates with decreasing amplitude converging after 12~fs to its final value. This decay arises from the reduced overlap of the superimposed molecular orbitals, which become soon located on the two different nuclei during the dissociation process. Consequently, the last term of eq.~\ref{eq:Dens} vanishes and hence the dynamics of the electronic linear combination. The probability $P_\mathrm{C^+}(t)$ is finally given by the prepared population distribution of the coupled reaction channels leading either to C$^+$ or O$^+$. This ratio is steered very precisely by the CEP modulating the relative intensities between half-cycles and is only significant in ultrashort laser pulses. When multiple half-cycles are taken into account the mechanism of the observed control stays the same. A shift of the CEP by $\pi$ while keeping the molecular orientation leads to a different result (green curve in fig.~\ref{fig:QD}(c)). Changing the orientation of the molecule by 180° is equal to shifting the CEP by $\pi$ as the transition dipole moment changes the sign. Thus the CEP dependent asymmetry in the dissociation step can be calculated by $P_\mathrm{C^+}(t)$ for two CEP values shifted by $\pi$. As shown in fig.~\ref{fig:QD}(c), the probability of measuring a C$^+$ fragment upon the break up of the molecule can be changed by 5\% through the CEP.

We presented experimental and theoretical results on the steering of electrons in a multi-electron system. As possible mechanisms for the observed CEP-control of the directional fragment emission in the dissociative ionization of CO we have discussed contributions from the ionization step as well as from the laser induced dynamics during the dissociation. The current experimental data does not allow to clearly distinguish the individual contributions. Further studies are underway that aim into this direction.

\begin{acknowledgments}
We acknowledge F. Krausz for his support and we thank Z. Chen and C. D. Lin for their help with the CEP calibration. The work of M.J.J.V. is part of the research program of FOM (financially supported by NWO). We are grateful for support by the DFG via the Emmy-Noether program and the Cluster of Excellence: Munich Center for Advanced Photonics.
\end{acknowledgments}
%

%
\bibliographystyle{apsrev}
\bibliography{Literature}
\end{document}